\documentclass[twocolumn,showpacs,preprintnumbers,prl,aps,amssymb,superscriptaddress]{revtex4}
\usepackage{amsmath}
\usepackage{stmaryrd}
\usepackage{txfonts}
\usepackage{amssymb}
\usepackage{mathrsfs}
\usepackage{graphicx}
\usepackage{dcolumn}
\usepackage{bm}
\usepackage{epsfig}
\usepackage{color}
\usepackage{hyperref}
\begin{document}
\title{Triplet superconductivity in ferromagnets due to magnon exchange}
	\author{Lev Bulaevskii}
	\affiliation{International Institute of Physics, Universidade Federal do Rio Grande do Norte, Campus Universit\'{a}rio, Lagoa Nova, Natal-RN 59078-970, Brazil}
	\affiliation{Department of Physics, University of Oregon, Eugene, USA}
	\author{Ronivon Eneias}
	\affiliation{International Institute of Physics, Universidade Federal do Rio Grande do Norte, Campus Universit\'{a}rio, Lagoa Nova, Natal-RN 59078-970, Brazil}
	\author{Alvaro Ferraz}
	\affiliation{International Institute of Physics, Universidade Federal do Rio Grande do Norte, Campus Universit\'{a}rio, Lagoa Nova, Natal-RN 59078-970, Brazil}
	\affiliation{Departamento de F\'{i}sica Te\'{o}rica e Experimental, Universidade Federal do Rio Grande do Norte, Campus Universit\'ario, Lagoa Nova, Natal-RN 59078-970, Brazil}
	
	\begin{abstract}
		
		We consider the superconducting pairing induced by spin waves exchange in a ferromagnet with both conduction and localized electrons, the latter being described as spins. We use the microscopic Eliashberg theory to describe the pairing of conducting electrons and the RPA approach to treat the localized spins assuming an exchange coupling between the conducting electrons and spins.  In the framework of non relativistic Hamiltonian twe found that he spin wave exchange results in equal spin electron pairing described by the two components  of the order parameter, $\Delta^{\uparrow}$ (both spins up) and  $\Delta^{\downarrow}$ (both spins down).
		Due to the conservation of total spin projection on the axis of the spontaneous ferromagnetic moment, the spin wave exchange at low temperatures includes an emission of magnons and an absorption of thermal magnons by the conduction electrons. 
		The absorption and emission processes depend differently on the temperature, with the absorption being progressively suppressed as the temperature drops. As a result,  the superconducting pairing exists only if the electron-spin wave exchange parameter $g$ exceeds some critical value $g_c$. At $g>g_c$
		pairing vanishes if the temperature drops below the lowest point $T_{cl}$ or increases above the upper critical point $T_{ch} \approx T_m$
		(the Curie temperature) where the spin waves cease to exist. This behavior inherent to the spin carrying glue is in an obvious disagreement with the results of conventional BCS approach which assumes that the effective electron-electron attraction is simply proportional to the static magnetic susceptibility. 
	\end{abstract}
	

	\maketitle

	\date{\today}


	\section {Introduction}
	 Eliashberg \cite{Eliashberg,AGD} and Scalapino, Schrieffer and Wilkins \cite{SSW} developed the microscopic approach  for  the BCS phonon induced Cooper pairing in standard superconductors like Nb and Pb. This electron-phonon model explicitly introduces phonons (weak ion distortions from their equilibrium positions in the crystal lattice) under the effect of electrons resulting in the effective indirect electron attraction. In the framework of this model involvement of phonons may be checked by tunneling measurements. Namely, in this model, the origin of the pairing mechanism is imprinted into the superconducting order parameter via its frequency dependence on the density of states of phonons working as a glue. Such a dependence may be revealed by the measurements of the I-V characteristics of the tunneling between the superconductor and the normal metal. Then peaks in the  I(V) should coincide with those observed in neutron scattering experiments. The experimental results indeed confirm the phonon mechanism and the dynamic nature of the glue in superconductors such as Nb, Al, In, Pb. What is more, McMillan \cite{McMillan} has shown that tunneling data provides all the information needed to calculate the superconducting critical temperature $T_c$ which can then be compared with the experimental $T_c$ to check the consistency of the model. After all these theoretical results have been confirmed by their corresponding measurements, no doubt remained about the phonon origin of the glue in standard superconductors with singlet electron pairing. 
	
	However, there is doubt that the phonon mechanism of Cooper pairing can explain the behavior of a broad family of high-temperature superconductors, although a predictions of the BCS phenomenological approach are still used to describe some of their experimental properties. It would be instructive to develop at least yet another microscopic model to see whether or not the predictions of the BCS approach are universal in describing superconducting properties, at least on the phenomenological level, in other scenarios.
	
	Recently, anomalous superconducting properties were observed in the U-based compounds UGe$_2$, URhGe and UCoGe which are quite different from the well known singlet superconductors, see \cite{Aoki1,Aoki2,Mineev} First, superconductivity was observed in their ferromagnetic state below the Curie temperature $T_m$ (observed superconducting critical temperature $T_c<T_m$). The manifestation of pairing takes place deep in the ferromagnetic state, with the internal exchange field suppressing the Cooper pairing of electrons with opposite spins in an indication that, in those compounds, the pairing may be in the triplet channel rather than in the singlet one. The observation of an upper critical field well above the paramagnetic limit for singlet superconductors \cite{Aoki2} also favors triplet pairing, although an enhanced spin-orbital coupling in materials with heavy elements like U may significantly weaken the destructive effects of the internal exchange field and of the applied magnetic field on superconducting singlet pairing. If there is indeed triplet pairing, one can exclude the phonon mechanism because it provides stronger attraction in the singlet channel. If phonons cannot be effective for pairing in ferromagnetic phases of U-based compounds, the only probable glue for pairing is the exchange via spin waves. This logic led McHale, Hattori \cite{Hale,Hattori} and Mineev\cite{Mineev} to consider respectively spin waves and magnetic fluctuations near $T_m$ as the glue mechanism for the superconducting pairing at $T_c$ just below $T_m$. 
	
	The replacement of phonon glue by the spin wave glue looks to be natural move which should not lead to drastic modification of superconducting properties. However, there is important difference between these types of glue. Namely, the spin waves (magnons) carry a projection of spin on the direction of ferromagnetic moment and the non relativistic approach the projection of total spin of both conduction and localized electrons is preserved. Thus the exchange by magnons is restricted by this conservation law, while such restriction does not exist for exchange by spin less phonons. One can thus anticipate that the magnon driven superconductors may not follow the BCS scenario. In the following, we will find that this is really the case in the non relativistic approximation.  
	
	In this paper, we consider the non relativistic Hamiltonian with RKKY interaction of conducting electrons and localized spins. We use the microscopic Eliashberg theory to treat the triplet superconductivity induced by the spin waves (magnons) exchange in the presence of ferromagnetic ordering. To describe the superconductor at all temperatures we use the Tyablikov \cite{Tyab} random phase approximation, while for the electrons and their interaction with spin waves we use the Gor'kov-Nambu formalism. We show that the spin wave exchange mechanism results in the equal spin pairing described by two component order parameter. We show that it is the spin conservation at electron-magnon scattering which determines the coupling of these two components and the resulting specific low-temperature behavior of triplet superconductors. 
	We find the corresponding phase diagram which is indeed drastically different from that for the standard phonon based BCS superconductors.  
	We also show that as in the phonon driven superconductors, the tunneling measurements in ferromagnet superconductors reveal the origin of the dynamics of the glue, i.e. they show the peculiarities of the magnon spectrum in this case.   
	
	\section {The Hamiltonian}
	
	We consider a system of conduction  electrons in a broad $s$-band with a two dimensional dispersion $\epsilon({\bf k})$ and localized spins $S$ describing the  spin degrees of localized electrons in the narrow $f$ or $d$-band positioned below the Fermi level and occupied only partially due to the strong Coulomb electron repulsion in that band. We assume that the localized spins interact with each other by means of the direct spin-spin Heisenberg exchange and  also via the RKKY interaction with the conduction electrons. In the non relativistic approach (ignoring spin-orbit coupling) the system is described by the model Hamiltonian:
	\begin{eqnarray}
	&&{\cal H}={\cal H}_{e}+{\cal H}_s+{\cal H}_{int}, \label{1}\\ 
	&& {\cal H}_{e}=
	\int d{\bf r}\hat{\psi}_{\alpha}^+({\bf r})\left[\epsilon({\bf k})-\epsilon_F\right]\hat{\psi}_{\beta}({\bf r}),\label{sec}\\
	&&{\cal H}_s=
	\sum_{m,n}[J_{{\rm d},{\bf n},{\bf m}}(S_{\bf m}^xS_{\bf n}^x+S_{\bf m}^yS_{\bf n}^y)+J'_{{\rm d},{\bf n},{\bf m}}S_{\bf m}^zS_{\bf n}^z]+\nonumber\\
	&&~~~~~~~~~~~\frac{1}{2}K(S_{x{\bf m}}^2+S_{y{\bf m}}^2), \label{3} \\
	&&{\cal H}_{int}=\int d{\bf r}\sum_{{\bf n},i,\alpha,\beta}\left[h_{ex}({\bf r},{\bf n})\psi_{\alpha}^+({\bf r})\sigma_{\alpha\beta}^iS_{\bf n}^i\psi_{\beta}({\bf r})+c.c.\right]. \label{ex}
	\end{eqnarray}
	Here $\epsilon({\bf k})=k^2/2m$ is the electron dispersion with the effective mass $m$, the electron momentum is ${\bf k}=(k_x=k\cos\phi, k_y=k\sin\phi)$, where $\phi$ is the polar angle in the $x,y$ plane and the $\sigma^i$ are the Pauli matrices. The indices $\alpha$ and $\beta$ denote spin-up and down states of the conduction electrons. The parameters $J_{\rm d}$ and $J'_{\rm d}$ describe direct exchange between localized spins, while the parameter $K$ describes the magnetic anisotropy energy \cite{Woolsey}. The exchange interaction between the conduction electrons and  localized spins which is characterized by the parameter $h_{\rm ex}$, results in an indirect RKKY interaction of the localized spins.  We do not account here for the orbital effect of the internal magnetic field induced by localized spins. It may be accounted for by the renormalization of the superconducting critical temperature.
	We do not account also for the Coulomb repulsion between the conduction electrons because, as we will show next, the magnon exchange results in Cooper triplet spin pairing of electrons with aligned spins only. Due to the Pauli principle, the repulsion of such electrons is significantly weakened. 
	
	We note that the simple Hamiltonian used here does not pretend to describe the real complex situation of the $f$-electrons in U-compounds, with partially occupied $f$-band representing both localized and itinerant electrons (see, Refs.~\onlinecite{Hale},  \onlinecite{Hattori} and references therein). Nevertheless, this minimal Hamiltonian model allows us to account for low energy electron and localized spin degrees of freedom. Another important point is that it conserves the $z$-projection of the total spin of the system. The interplay of electron and spin degrees of freedom in the framework of spin conservation law results in the main peculiar property of this triplet superconductivity -- a two-component structure of order parameters for triplet pairing, the spin preserved coupling of these two components and its effect on the phase diagram of triplet superconductors.	
	
	We assume that the localized spins are ordered ferromagnetically along the axis $z$, and $S_z=\langle S_{z,{\bf n}}\rangle_{\bf n}$ is the ferromagnet order parameter. In this case the spin waves (magnons) are collective small deviations 
	of the magnetization along the $x$ and $y$ axis and they are described by the spin operators $S_{\bf n}^{\pm}=S_{x,{\bf n}}\pm iS_{y,{\bf n}}$, acting  as the creation and the annihilation of spin waves, $|S_{x,{\bf n}}|, |S_{y,{\bf n}}|\ll S_{z}$. In the momentum representation we use the operators $S_{\bf q}^{\pm}=S_{x,{\bf q}}\pm iS_{y,{\bf q}}$ where ${\bf q}$ is the  spin wave momentum.
	The commutation relations of the spin operators ${\bf S}_{{\bf n}}^{\pm}$ obey the conventional SU(2) algebra:
	\begin{eqnarray}
	S^{+}S^{-}-S^{-}S^{+}=2S_z.
	\end{eqnarray}
	In the spin-wave random phase approximation (RPA) \cite{Tyab} we replace the operator $S_{z,{\bf n}}$ in the right hand side of this equation by its thermal average $\langle S_z\rangle_T$.
	The spin operators we replace by $S_{\bf n}^{\pm}=(S_{x,{\bf n}}\pm iS_{y,{\bf n}})/(2\langle S_z\rangle_T)$. They obey boson commutation relation and play the role of creation and annihilation operator of spin waves at the site ${\bf n}$. In the momentum representation the operators $S_{\bf q}^{\pm}=S_{x,{\bf q}}\pm iS_{y,{\bf q}}$ are the creation and annihilation operators of the spin wave (magnon) with the momentum ${\bf q}$ and with the dispersion
	\begin{eqnarray}
	\epsilon_s({\bf q})=\langle S_z \rangle(K+v_s^2{\bf q}^2), \label{dis}
	\end{eqnarray}
	where $v_s$ is the magnon velocity so that $v_s^2=cT_m$, with the numerical coefficient $c$  of the order unity. Indeed, the Hamiltonian Eq.~(\ref{3}) results in the equations of motion for spin wave operators  in the Fourier representation with respect to coordinates and imaginary time $\tau$:
	\begin{eqnarray}
	\dot{S}_{{\bf q}}^{\pm }=\pm\langle S_z \rangle\epsilon_s({\bf q})S_{{\bf q}}^{\pm}. \label{dis}
	\end{eqnarray}
	Now magnons play the same role of bosonic excitations as phonons and thus 
	in the spin wave RPA approximation we can use the Wick's theorem to develop the perturbation theory with respect to the electron-magnon interaction. 
	
		Account for the spin-orbit coupling results in more general form of the interaction part of the Hamiltonian
	\begin{eqnarray}
	&&{\cal H}_{int}=\nonumber\\
	&&\int d{\bf r}\sum_{{\bf n},\gamma,\kappa,\alpha,\beta}\left[h_{ex}^{\gamma\kappa}({\bf r},{\bf n})\psi_{\alpha}^+({\bf r})\sigma_{\alpha\beta}^{\gamma}S_{\bf n}^{\kappa}\psi_{\beta}({\bf r})+c.c.\right]. 
	\end{eqnarray}
	The terms non diagonal in the indices $\gamma$ and $\kappa$ are absent in non relativistic approximation.  and their relation to the diagonal ones 
	may be estimated as $\delta g/g$, where $g$ is the electron gyromagnetic ratio and $\delta g$ is its deviation from 2. For metals Fe, ($Z=26$) Cr ($Z=24)$ 
	and Mn ($Z=25)$ the ratio $\delta g/g$ is 10$^{-3}, 10^{-2}$ and $10^{-3}$, respectively. \cite{Moriya} Taking into account that this ratio depends on the atomic number $Z$ as $Z^4$ we  see that $\delta g/g\approx 3(10^{-8}-10^{-9})Z^4$. Thus non relativistic Hamiltonian is a good  approach for $d$-wave Fe, Mn, Cr and rare earth metals. For U compounds 
	relativistic part of the Hamiltonian may be important and our results may be invalid for them.

	Though the description of the phonon glue and of magnon glue look very similar, an important difference exists between their interactions with electrons. Phonons do not carry spins, while magnons do carry $\pm 1$ spin projections for $S_{\bf q}^{\pm}$, respectively. As a result, the effective electron-electron interaction becomes spin-dependent with on-site vertexes proportional to $\sum_i\sigma_iS_i=(\sigma^+S^-+\sigma^-S^+)\langle S_z\rangle$. Due to spin conservation for non-relativistic electron-spin interaction Hamiltonian, the complete cycle of magnon exchange should include both the emission and the absorption of magnons. In view of that, the effective pairing electron-electron interaction ${\cal E}_{ee}$ is proportional to the product of those two matrix elements. In other words, ${\cal E}_{ee}$ {\it is not} proportional to the static magnetic susceptibility as was assumed in Refs.~\onlinecite{Mineev,Hattori,Kirkpatric} in the BCS approach for ferromagnetic superconductors. 
	Consequently, ${\cal E}_{ee}$ itself becomes strongly temperature dependent because the absorption of magnon is strongly suppressed at low temperatures. Such a limitation is absent in the case of a phonon glue.
	
	At low temperatures  $T\rightarrow 0$ the RPA replacement gives exact results because the amplitudes of the spin waves are negligible. At nonzero temperature in the RPA approach one needs to find $\langle S_z\rangle_T$ by solving appropriate self-consistent equations. For spin $S=1$, which will be considered later on, it has the form
	\begin{eqnarray}
	\frac{1}{\langle S_z \rangle}=\frac{1}{\pi}\int_0^{\pi}d{\bf q}\coth\frac{\epsilon_s({\bf q})}{2T}.
	\end{eqnarray}
	
	The magnon correlation function in the Fourier representation with respect to coordinate and imaginary time $\tau$ is, in turn,
	\begin{eqnarray}
	\langle S_{\bf n}^{\mp}(\tau)S_{\bf m}^{\pm}(\tau')\rangle_{\nu_n,{\bf q}}= \langle S\rangle_T[i\nu_n\mp \epsilon_s({\bf q})]^{-1} , \ \ \ \nu_n=2\pi n.
	\label{magnon}
	\end{eqnarray}
	Note, that in the RPA approach, the frequencies of all magnetic excitations drop proportionally to $\langle S_z \rangle_T$ as $T$ approaches $T_m$ and vanish at $T>T_m$. Thus this approach does not account for strong magnetic fluctuation near the Curie temperature.
	
	In our calculations, we will consider the spin dynamics unaffected by superconducting pairing. This approach is valid because the free energy of the magnetic system is much larger than the superconducting condensation energy at all temperatures, except at the very narrow region near $T_m$. Indeed, the free energy of localized spins is of the order of $T_m$ per spin and it drops as $(T_m-T)^2/T_m$ as $T$ approaches $T_m$. In contrast, the energy of the superconducting pairing is of the order of $(T_c-T)^2N(0)$ and $T_c\leq T_m\ll \epsilon_F$. Here $N(0)$ is the electron density of states per spin and $N(0)\approx 1/\epsilon_F$, with $\epsilon_F$ being the electron Fermi energy, Thus only at temperatures in which $(T_m-T)/(T_m-T_c)\approx (T_m/\epsilon_F)^{1/2}$, magnetism and superconducting pairing strongly affect each other.
	
	The Curie temperature $T_m$ of ferromagnetic ordering is determined by the sum of the RKKY interaction
	and of the direct spin coupling  contribution $T_d=\sum_mJ'_{{\rm d},{\bf n},{\bf m}}$, i.e. $T_m=c_1J_{\rm ex}+c_2J_{\rm d}$, where $J_{\rm ex}\approx h_{\rm ex}^2N(0)$,  and $c_1,c_2$ being numerical coefficients.

	\section {Nonunitary triplet order parameter}
	
	The description of the triplet superconducting state is quite different from that of a singlet state because, in the first place, the order parameter should be characterized by a vector rather than a scalar. 
	For spin triplet pairing the superconducting order parameter $\hat{\Delta}$ is characterized by the vector ${\bf d}$ as  
	\begin{eqnarray}
	&&\hat{\Delta}({\bf r},{\bf k})=id_i({\bf r},{\bf k})\sigma_i\sigma_y=\sum_iD_i({\bf r},{\bf k})\sigma_i,\\
	&&D_0=id_y, \ \ \ D_x=d_z, \ \ \ D_y=0 \ \ \ D_z=-d_x,
	\end{eqnarray}
	were summation over repeated indices $i=0,x,y,z$ is assumed. Here and in the following we denote $\sigma_0=
	\hat{1}$. The order parameter depends on the center of mass ${\bf r}$ and on the momentum ${\bf k}$ of the Cooper pair.
	It is odd in both coordinates and in momenta.
	In the momentum space the vector ${\bf d}$ is related to the amplitudes of the spin-up, the spin-down and the zero-spin projections of the superconducting order parameter $\Delta^{\uparrow}, \Delta^{\downarrow}, \Delta^{0}$ as follows 
	\cite{Mineev}
	\begin{eqnarray}
	&&d_x=(1/2)(-\Delta^{\uparrow}+\Delta^{\downarrow}), \\
	&&d_y=-(i/2)(\Delta^{\uparrow}+\Delta^{\downarrow}),\label{Delta}\\
	&&d_z=\Delta_0.
	\end{eqnarray}
	Here $\Delta^{\uparrow}$ is the amplitude of the state $|\uparrow\uparrow\rangle$,  $\Delta^{\downarrow}$ is the amplitude of the state $|\downarrow\downarrow\rangle$, while $\Delta_0$ is the amplitude of the state $|\uparrow\downarrow\rangle+|\downarrow\uparrow\rangle$. We note that while the superconducting state characterized by the parameter $d_z$ corresponds to an electron pair with spin oriented in the direction perpendicular to the $z$-axis, states with parameters $(d_x,d_y)$ correspond to electron system of at least two up-spins or two down-spins along the $z$-axis. Exchange by spin waves  does not lead  to the interaction of electrons with all spins aligned because the spin wave carries  unity spin and thus the spins of two interacting electrons should be opposite to each other. Hence, pairing in the system of two or more electrons with aligned spins does not exists. In contrast, in the systems of four or more electrons with different spins the ordering characterized by the parameters $(d_x,d_y)$ is possible, due to the presence of attraction between electrons with opposite spins. 
	
	For triplet pairing the dependence of the amplitudes  $d_x,d_y$ on momentum ${\bf k}$ of Cooper pair electrons should be odd. The function $d_z$ should be also odd in ${\bf k}$, but we will see that $d_z=0$ in the magnon exchange model. 
	Thus the two functions $d_{x,y}$ fully characterize the triplet order parameter in the magnon exchange model.
	Singlet pairing is characterized by the scalar $d_0({\bf k})$ and $\hat{\Delta}=id_0\sigma_y$, with $D_y=id_0$.
	
	\section{Self-consistent equations for superconducting order parameters }
	
	We neglect the vertex corrections for the electron-magnon coupling, as was done for electron-phonon case, because we assume $\epsilon_s({\bf q})\ll\epsilon_F$.
	We consider two dimensional electron system with momenta ${\bf k}=k_F(\sin\phi,\cos\phi)$, 
	while change of electron momentum from ${\bf k}$ to ${\bf k'}$ at the electron-magnon scattering requires conservation of the momentum
	\begin{eqnarray}
	{\bf k}-{\bf k}'={\bf q}.
	\end{eqnarray}
	In the superconducting state we write down the electron Green's function in the Nambu representation as a  $2\times 2$ matrix  $\hat{G}(\omega,{\bf k})$, whose diagonal components $G_{11}$ and $G_{22}$ are the 
	conventional  Green's functions, while the off-diagonal elements $G_{12}$ and $G_{21}$ are the Gor'kov's functions $F$ and $F^+$ describing the pairing condensation. They are matrices in the 
	spin coordinates and in general case may be expand as a sum in the  $\sigma_i$ matrices:
	\begin{eqnarray}
	\hat{F^+}(\omega_n,{\bf k})=\sum_ia_i({\bf k},i\omega_{n})\sigma_i, \ \ \  i=0,x,y,z.  \label{exp}
	\end{eqnarray}
The Eliashberg equations for the Green's functions in the electron-phonon model in the Matsubara representation have the form (see Refs. \onlinecite{Eliashberg,AGD})
\begin{eqnarray}
&&[i\omega-\xi({\bf k})]G(\omega,{\bf k})=\\
&&1+g^2\int d{\bf q}G(\omega-\nu,{\bf k}-{\bf q})D(\nu,{\bf q})G(\omega,{\bf k})+\nonumber\\
&&g^2\int d{\bf q}F(\omega-\nu,\omega,{\bf k}-{\bf q})D(\nu,{\bf q})F^+(\omega,{\bf q}),\nonumber \\
&&[-i\omega-\xi({\bf k})]F^+(\omega,{\bf k})	= \label{second}\\
&&1+g^2\int {\bf q}G(\omega-\nu, {\bf k}-{\bf q})D(\nu,{\bf q})F^+(\omega,{\bf k})+\nonumber\\
&&g^2\int d{\bf q}F^+(\omega-\nu,{\bf k}-{\bf q})
D(\nu,{\bf q})G(\omega,{\bf k}),\nonumber
\end{eqnarray}
where $\omega$ denotes $\omega_n=2\pi(n+1/2)$, while $\xi({\bf k})$ is the electron energy accounted from the Fermi energy. Further, $D(\nu,{\bf k})=-\langle  T[\varphi(x_1)\varphi(x_2)]\rangle$ is the phonon Green function,
$\nu=2\pi n$ and $g$ is the electron-phonon coupling parameter. 

For electron-magnon model we need to replace the phonon Green function by the spin correlation functions Eq. (\ref{magnon}) and the couple parameter $g$ by 
$J(k,k')$ and account for spin conservation at the vertex, i.e. we replace 
\begin{eqnarray}
 &&g^2\sum_{\nu}\int d{\bf q}D(\nu,{\bf q})F^+(\omega-\nu,{\bf k}-{\bf q})\Rightarrow \\
 &&\sum_{\nu}\int d{\bf q} |J({\bf k},{\bf k}-{\bf q})|^2\times\\
 &&[S^{+-}(\nu,{\bf q})\sigma^-F^+(\omega-\nu,{\bf k}-{\bf q})\sigma^++\nonumber\\
 &&S^{-+}(\nu,{\bf q})\sigma^+F^+(\omega-\nu,{\bf k}-{\bf q})\sigma^-].\nonumber
 \end{eqnarray}
Neglecting the the renormalization of the electron Green function due to electron-magnon coupling, the electron diagonal Green's function in the absence of external magnetic field is
	\begin{eqnarray}
	 G_{0}(\omega,{\bf k})^{-1}=i\omega_n-\xi-h_z\sigma_z.
	\end{eqnarray}
	The exchange field $h_z$ splits the electron energy levels into two electron bands with spins up and down. At strong $h_z$ one of these bands becomes empty, while the other is filled with spins of corresponding kind. As was mentioned above, superconducting pairing is absent in this case because electrons of the same kind do not interact with each other by spin wave exchange.  At low $h_z$, electrons with both spin up and down are present close to the Fermi level and contribute to pairing. This case is discussed in the following.

 From Eq.~(\ref{second}), in the second order perturbation theory with respect to the electron-spin exchange interaction term in Eq.~(\ref{ex}), we obtain the self-consistent equation for the Gor'kov's Green function $F^+(\omega,{\bf k}$ at $T_c$:
	\begin{eqnarray}
	&&F^+(\omega_n,{\bf k})=\sum _ia_i\sigma_i= \label{RHS}\\
	&&-2T\sum_{{\bf k}',\omega_{n'}}|J({\bf k},{\bf k'})|^2
	G_0(-\omega_{n'},{\bf k'})G_0(\omega_{n'}{\bf k')}\times\nonumber\\
	&&[S^{+-}(\nu,{\bf q})\sigma^-F^+(\omega_{n'}-\nu,{\bf k}-{\bf q})\sigma^++\nonumber\\
	&&S^{-+}(\nu,{\bf q})\sigma^+F^+(\omega_{n'}-\nu,{\bf k}-{\bf q})\sigma^-]\nonumber
	\end{eqnarray}
	with the spin correlation functions given by Eq.~(\ref{magnon})
	and
	\begin{eqnarray}
	&&J_{{\bf k}{\bf k'}}=\int d{\bf r}h_{\rm ex}({\bf r})e^{i({\bf k}-{\bf k'})\cdot{\bf r}}. \label{last}
	\end{eqnarray}
	Here in the right side we introduced the magnon Green's functions $S_{ij}(i\nu_n,{\bf q})$. 
	\begin{eqnarray}
	&&{\cal S}_{+-}(\tau-\tau', {\bf n}-{\bf m})=\langle S_{\bf n}^+(\tau)S_{\bf m}^-(\tau')\rangle\label{sp2}.\\
	&&{\cal S}_{-+}(i\nu_m,{\bf q})=\int d\nu (i\nu_m-\nu)^{-1}\delta[\nu-\epsilon_s({\bf q})], \\
	&&{\cal S}_{+-}(i\nu_m,{\bf q})=\int d\nu (i\nu_m+\nu)^{-1}\delta[\nu-\epsilon_s({\bf q})].
	\end{eqnarray}
	
	First, we prove that self-consistency equation results in the relation $D_x=0$, i.e. $d_z=0$. For that we write the electron Green's function $\hat{G}$ 
	in the form Eq.~(\ref{exp}). Then in the right hand side of Eq.~(\ref{RHS}) we have that
	\begin{eqnarray}
	&&\sigma_+\sigma_{x,y}\sigma_-=\sigma_-\sigma_{x,y}\sigma_+=0, \label{sing}\\
	&&\sigma_{\pm}\sigma_0\sigma_{\mp}=2(\sigma_0\pm\sigma_z), \nonumber\\
	&&\sigma_{\pm}\sigma_z\sigma_{\mp}=\mp2(\sigma_0\pm\sigma_z).\nonumber
	\end{eqnarray} 
	Due to  the first relation above, the matrices $\sigma_x$ and $\sigma_y$ are absent in the right hand side of Eq.~(\ref{RHS}) and, consequently, in the left hand side they are absent too. 
	As a result, only the components  $d_x,d_y$ are nonzero. This means that the magnon exchange results in equal-spin triplet states
	$\hat{\Delta}=(\Delta^{\uparrow}|\uparrow\uparrow\rangle, \Delta^{\downarrow}|\downarrow\downarrow\rangle)$. The opposite statement is also true, namely if we have a superconductor with such $\hat{\Delta}$, the glue should carry spin $S_z=\pm 1$ because otherwise, the coupling of its component does not obey the spin conservation law. Note, that relations (\ref{sing}) exclude also singlet pairing induced by the magnon exchange mechanism.
	
To find the superconducting critical temperature of the transition from normal state to pairing state, $T_c$, and the structure of the superconducting order parameter at $T_c$ it is sufficient  to know $F^+$ at in the lowest order in the components of the order parameter, $d_x$ and $d_y$. 	Thus we write Eq.~(\ref{RHS}) in the form 
	\begin{eqnarray}
	&&F^+(\omega_n,{\bf k})=-2T\sum_{{\bf k'},\omega'}|J_{{\bf k}{\bf k'}}|^2\times\nonumber\\
	&&\left\{{\cal S}_{-+}(i\omega-i\omega',\phi-\phi') \Lambda\sigma_-(D_0\sigma_0+D_z\sigma_z)\sigma_++\nonumber\right.\\
	&&\left.{\cal S}_{+-}(i\omega-i\omega',\phi-\phi')\Lambda\sigma_{+}(D_0\sigma_0+D_z\sigma_z)\sigma_-\right\},\label{sim}\\
	&&\Lambda=G_0(-\omega,{\bf k'})G_0(\omega,{\bf k'})=\\
	&&\frac{\xi'^2+\omega^2+h_z^2-2\xi'h\sigma_z}{[(\xi'+h_z)^2+\omega^2][(\xi'-h_z)^2+\omega^2]}.
	\nonumber
	\end{eqnarray}
		
	Next we integrate over $\xi'$ and take into account that at $h_z\ll\epsilon_F$ we may put $h_z=0$.  Summation over Matsubara frequencies $\omega $ of Green's function $F^+(\omega,{\bf k})$  we replace
	by the integral over real frequencies $\omega'$
	as was done in (2.12) of Ref.~\onlinecite{SSW} for the superconductor 
	with electron-phonon coupling. We obtain the equation
	\begin{eqnarray}
	&&F^+(\omega_n,{\bf k})=-\sum_{{\bf k'}}|J_{{\bf k}{\bf k'}}|^2\int_{-\infty}^{+\infty}\frac{d\omega'}{\omega'}\nonumber\times\\
	&&\left\{\frac{\sigma_+(D_0\sigma_0+D_z\sigma_z)\sigma_-}{\omega'-(i\omega_n+\epsilon_s({\bf k}-{\bf k}'))}[f(\omega')+n(\epsilon_s)]\right.-
	\nonumber\\
	&&\left.\frac{\sigma_- (D_0\sigma_0+D_z\sigma_z)\sigma_+}{(\omega'-i\omega_n+\epsilon_s({\bf k}-{\bf k}'))}[f(-\omega')+n(\epsilon_s)]\right\}
	\label{FF}\\
	&&f(\omega)=[e^{\beta\omega}+1]^{-1}, \ \ \ \ n(\epsilon_s)=[e^{\beta\epsilon_s}-1]^{-1},
	\end{eqnarray}
	where $f(\omega)$ and $n(\omega)$ are the fermion and the boson distribution functions, and $\beta=1/T$. 
	Now the expression (\ref{FF})  for $F^+(\omega_n)$ can be analytically continued with respect to $i\omega_n$ to the real axis from the upper half-plane by replacing $i\omega_n$ by $\omega+i\delta$:
	\begin{eqnarray}
	&&F^+(\omega,\phi)=D_0\sigma_0+D_z\sigma_z=\nonumber\\
	&&-\int_0^{2\pi}d\phi'|J(\phi,\phi')|^2\int_{-\infty}^{+\infty}\frac{ d\omega'}{\omega'}\times\nonumber\\
	&&\left\{\frac{[{\rm Im}~(D_0-D_z)](\sigma_0+\sigma_z)}{\omega'-i\delta-\omega-\epsilon_s(\phi-\phi')}\left[\frac{1}{1+e^{\beta\omega'}}+\frac{1}{e^{\beta\epsilon_s}-1}\right]-\right. \nonumber\\
	&&\left.\frac{[{\rm Im}~(D_0+D_z)](\sigma_0-\sigma_z)}{\omega'-i\delta-\omega+\epsilon_s(\phi-\phi')}\left[\frac{1}{1+e^{-\beta\omega'}}+\frac{1}{e^{\beta\epsilon_s}-1}\right]\right\}. \label{rr}
	\end{eqnarray}
	We introduce notations
\begin{eqnarray}
	&&{\cal D}(\omega,\phi)=D_0(\omega, \phi)-D_z(\omega,\phi)=\Delta^{\downarrow}(\omega,\phi),\nonumber\\
	&&{\cal R}(\omega,\phi)=D_0(\omega, \phi)+D_z(\omega,\phi)=\Delta^{\uparrow}(\omega,\phi), \nonumber
	\end{eqnarray}
	Equating the coefficients in front of operators $\sigma_0$ and $\sigma_z$ in both sides of  Eq.~(\ref {rr}) we obtain two coupled equations connecting the order parameters $\Delta^{\downarrow}$ and $ \Delta^{\uparrow}$ at the temperature of the second order phase transition $T_c$.  This is the temperature of vanishing superconducting order parameter at cooling and we will find in the following that so found critical temperature decreases with $g$. Here we introduce the dimensionless parameter of electron-magnon coupling $g=2|J_{{\bf k}{\bf k'}}S|^2N(0)/T_m$. The first equation is
	\begin{eqnarray}
	&&{\rm Im}~\Delta^{\downarrow}(\omega,\phi)
	=\nonumber\\
	&&4g\int_0^{2\pi}d\phi'\int_{-\infty}^{+\infty}\frac{ d\omega'}{\omega'}{[\rm Im}~\Delta^{\uparrow}(\omega',\phi')]\times\nonumber\\
	&&\left[\frac{1}{1+e^{-\beta\omega'}}+\frac{1}{e^{\beta\epsilon_s(\phi-\phi')}-1}\right]\delta[\omega'-\omega+\epsilon_s(\phi-\phi')].\label{Main1}
	\end{eqnarray}
	This equation describes the creation of a magnon with energy $\epsilon_s$ by an electron which changes its spin by -1, while its energy goes from $\omega'$ to $\omega'-\epsilon_s$. The second equation
	\begin{eqnarray}
	&&[{\rm Im}~\Delta^{\uparrow}(\omega,\phi)]=\nonumber\\
	&&-4g\int_0^{2\pi}d\phi'
	\int_{-\infty}^{+\infty}\frac{ d\omega'}{\omega'} [{\rm Im}~\Delta^{\downarrow}(\omega',\phi')]\times\nonumber\\
	&&\left[\frac{1}{1+e^{\beta\omega'}}+\frac{1}{e^{\beta\epsilon_s(\phi-\phi')}-1}\right]\delta[\omega'-\omega-\epsilon_s(\phi-\phi')], \label{Main2}
	\end{eqnarray}
	describes the absorption of a magnon with energy $\epsilon_s$ by the electron which changes its spin by +1 and suffers the energy change from $\omega'$ to $\omega+\epsilon_s$. 
	Here $\epsilon_s(\phi-\phi')=\kappa+\gamma^2[1-\cos(\phi-\phi')]$ with $\gamma=2ak_F$ and $a$ is the spacing between localized spins, while $\kappa=K/T_m$.
	Writing the spin wave energy in this form we assume that $ak_F\lesssim 1$. In the following, we take $\gamma=1$. We also express all quantities with the dimension of energy in the units of $T_m$. 	
	
	Now we should account for the space dependence of the order parameters inherent to the triplet pairing. It should be odd in space variables because it is even in spin variable.
	Namely, the triplet superconducting order parameters $d_x(x,y)$ and $d_y(x,y)$ should change sign at the replacement $x\rightarrow -x$ and $y\rightarrow -y$.
	Thus both order parameters should change sign under the transformation 
	$\phi\rightarrow \phi+\pi$. Note, that for dispersion less magnons  the triplet pair order parameter vanishes because it must be odd in coordinate dependence (as opposed to the singlet ordering). To account for the change of sign at the  transformation $\phi\rightarrow \phi+\pi$, we expand 
	$\Delta^{\downarrow,\uparrow}(\omega,\phi)$ in a Fourier series with respect to the angle $\phi$:
	\begin{eqnarray}
	&&[{\rm Im}~\Delta^{\downarrow}(\omega, \phi)]=\sum_{n=1}\Delta^{\downarrow}_n(\omega)\sin[(2n-1)\phi]\label{an},\\
	&&[{\rm Im}~\Delta^{\uparrow}(\omega, \phi)]=\sum_{n=1}\Delta^{\uparrow}_n(\omega)\sin[(2n-1)\phi]\label{bn}.
	\end{eqnarray}
		We perform the integration over angles in right hand sides of Eqs. (\ref{Main1}) and (\ref{Main2}) by changing the variables in the integral. We get
	\begin{eqnarray}
	&&\sin\phi\sin\phi'=(\cos\alpha-\cos\beta)/2, \\
	&&\int_0^{2\pi}d\phi\int_0^{2\pi}d\phi'f(\phi-\phi',\phi+\phi')=\nonumber\\
	&&\frac{1}{2}\int_{-2\pi}^{2\pi}d\alpha\int_\alpha^{4\pi-\alpha}d\beta f(\alpha,\beta),
	\end{eqnarray}
	where $\alpha=\phi-\phi'$, while $\beta=\phi+\phi'$. 
	Integral  $\int_{\alpha}^{4\pi-\alpha}d\beta\cos\beta$  gives $2\sin\alpha$. Subsequent integration over $\alpha$ in the interval $(-2\pi,2\pi)$ 
	with a function even in $\alpha$ gives 0. 
	Then we integrate over angles and note that the states with different $n$ do not mix in this linear approximation. Thus the minimum $g$ at a given temperature, which we will call $T_{cl}(g)$, comes from the coupled equations for the least nonuniform state $n=1$. Hence, we obtain finally the pair of the integral equations 
	\begin{eqnarray}
	&&[{\rm Im}~\Delta^{\downarrow}(\omega)]=8g\int_{-\infty}^{+\infty} \frac{d\omega'}{\omega'}\Delta^{\uparrow}(\omega')\nonumber\times \\
	&&\frac{\omega'-\omega+\tilde{\kappa}}{[1-(\omega'-\omega+\tilde{\kappa})^2]^{1/2}}\left[\frac{1}{e^{-\beta\omega'}+1}+\frac{1}{e^{\beta(\omega-\omega')}-1}\right].
	\end{eqnarray}
	\begin{eqnarray}
	&&[{\rm Im}~\Delta^{\uparrow}(\omega)]=8g\int_{-\infty}^{+\infty} \frac{d\omega'}{\omega'}\Delta^{\downarrow}(\omega')\nonumber\times\\
	&&\frac{\omega'-\omega-\tilde{\kappa}}{[1-(\omega'-\omega-\tilde{\kappa})^2]^{1/2}}\left[\frac{1}{e^{\beta\omega'}+1}+\frac{1}{e^{-\beta(\omega-\omega')}-1}\right].
	\end{eqnarray}
	Here and in the following, the integration over $\omega'$ is  limited additionally by the condition that the expression under the square root should be positive definite. 
	Eliminating  $\Delta^{\uparrow}(\omega')$ in Eq.~(\ref{Main1}) we obtain a single equation for  $\Delta^{\downarrow}(\omega)$:
	
	\begin{eqnarray}
	&&\Delta^{\downarrow}(\omega)=-
	64g^2\int_{-\infty}^{+\infty}\frac{d\omega''}{\omega''}\Delta^{\downarrow}(\omega'')K(\omega,\omega''),\label{AK}\\
	&&K(\omega,\omega'')=\int_{-\infty}^{+\infty}\frac{d\omega'}{\omega'}\frac{\omega'-\omega+\tilde{\kappa}}{\left[1-\left(\omega'-\omega+\tilde{\kappa} \right)^2\right]^{1/2}}\times\nonumber
	\\
	&&\frac{\omega'-\omega''+\tilde{\kappa}}{\left[1-\left(\omega'-\omega''+\tilde{\kappa}\right)^2\right]^{1/2}}\left[\frac{1}{e^{-\beta\omega'}+1}+\frac{1}{e^{\beta(\omega-
			\omega')}-1}\right]\times\nonumber\\
	&&\left[\frac{1}{e^{\beta\omega''}+1}+\frac{1}{e^{\beta(\omega''-
			\omega')}-1}\right]. \label{KK}
	\end{eqnarray}
	
	Note, that two absorptions and two emissions of spin waves are needed to accomplish the cycle $\Delta^{\uparrow}\Rightarrow\Delta^{\downarrow}\Rightarrow\Delta^{\uparrow}$ for the Cooper pair. 	Both transitions are necessary because neither state $\Delta^{\uparrow}$ nor $\Delta^{\downarrow}$ can exist without each other since the spin wave carries spin 1 and thus the attraction of electrons with the same spins due to spin wave exchange is completely absent. 
	
	The temperature affects differently the rates of these two transformations $\Delta^{\uparrow}\Rightarrow\Delta^{\downarrow}\Rightarrow\Delta^{\uparrow}$. According to Eqs.~(\ref{AK}) and (\ref{KK}) the probability of the
	emission and subsequent absorption process
	are described by the factor 
	\begin{eqnarray}
	[f(\omega)+n(\epsilon_s)] [f(-\omega)+n(\epsilon_s)]=\frac{1}{4}\left[\cosh^{-2}\frac{\beta\omega}{2}+\sinh^{-2}\frac{\beta\epsilon_s}{2}\right].\nonumber\\
	\label{te}
	\end{eqnarray} 
	This factor, describing all temperature dependence of the triplet order parameter, drops exponentially as the temperature drops well below the anisotropy gap in the spin wave energy (if the gap is absent, the factor drops as power law with $T$). 
	This means that $T_{cl}$ drops as the coupling parameter $g$ increases. Such a tendency is opposite to the dependence $T_c$ in singlet phonon-mediated superconducting pairing described by a single order parameter in the BCS approach. In the latter case, the temperature dependence of the order parameter is described by the factor 
	$f(-\omega/T)-f(\omega/T)=\tanh(\omega/2T)$ at any $T$ because absorption of the phonon is not needed to exchange by phonons.
	
	\section{Numerical calculation}
	As we have integrated over angles we do not take care anymore of the angular dependence of the order parameter. Thus it suffices to solve the homogeneous integral equations ~(\ref{AK}), (\ref{KK})   for the functions of a single variable $\omega$.  For that, we convert the integral equation into an algebraic eigenvalue problem of the linearized Eliashberg equations.
	The first step is to convert the integrations over $\omega''$,  into a summation by using some quadrature scheme. Here we use the Gauss-Legendre quadrature $\int_{a}^{b}f(t)dt=\sum_{i=1}^{N}w_if(x_i)$ one of the many Gaussian-type quadrature formulas, where the $x_i$'s, also known as the abscissa, are the zeros of the Legendre polynomial - the integrand is evaluated at these points; and $w_i$ are the weights. More details of the quadrature method can be found in \cite{GW}.

	Thus, we construct the one dimensional matrix ${\cal A}_{nn''}$ corresponding to $K(\omega_n,\omega_{n''})$ to solve the equation ${\bf 1}-\lambda\hat{{\cal A}}=0$ by following a  standard diagonalization procedure for the matrix ${\cal A}_{nn''}$.
	
	After solving the equation  ${\bf 1}-\lambda\hat{{\cal A}}=0$ we find eigenvalues of corresponding matrix  at given temperature $T$. The maximum eigenvalue $g$ defines temperature $T_{cl}$, as well as the corresponding eigenvectors  $\Delta^{\downarrow}(\omega)$ and $\Delta^{\uparrow}(\omega)$. We will summarize our results in the  next session.

	\section{Boundaries of triplet superconducting phase in ferromagnet}

	The temperature at which  $\Delta^{\downarrow}(\omega)$ becomes nonzero determines the lower superconducting critical temperature $T_{cl}$ for the second order phase transition from the normal ferromagnetic state to the triplet superconducting state.
	The dependencies of $g(T_{cl})$ for $\kappa=$0.1 and $\kappa=0.2$ are shown in Fig.~1.
		\begin{figure}
		\psfig{figure=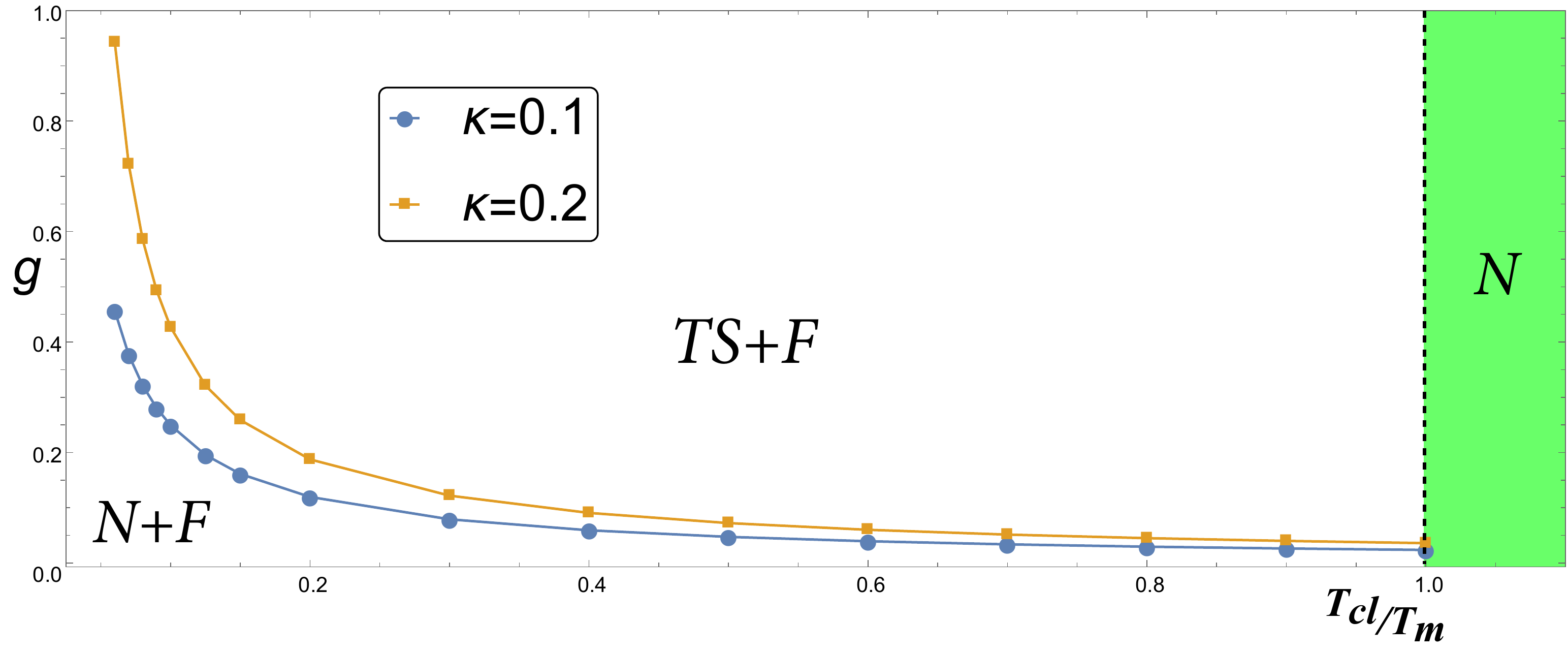,width=\columnwidth}
		\caption{\label{f2}(color online) The phase diagram of the ferromagnet in the plane the temperature $T_{cl}$ and the the strength $g$ of the exchange coupling of localized spins and conducting electrons. As the temperature increases from $T=0$, the superconducting triplet pairing establishes itself due to the exchange of electrons with spin waves at the second order transition line $T_{cl}(g,\kappa)$.  The descending line $g(T_{cl})$ is a direct consequence of the spin conservation at the magnon exchange which is effective only in the presence of thermal magnons. Note, that this drop in $g$ vs. $T_{c1}$ becomes more pronounced as the anisotropy parameter and thus gap in the spin wave spectrum increase.
			Warming up the superconductor  results in the destruction of spin waves and the Cooper pairs as the temperature approaches the Curie temperature $T_m$. This change from triplet superconducting phase to normal phase is signaled by a first order phase transition, its exact position may be found if magnetic fluctuations would be accounted for.}
	\end{figure}
	
	 In accordance with the previous discussion and in accordance with Eq.~(\ref{te}), $T_{cl}$ is a descending function of $g$, which is opposite to the behavior of $T_c(g)$ in phonon based superconductors. Note, however, that in our case $T_{cl}$ is the transition temperature from a normal ferromagnetic phase NF to a superconducting ferromagnetic phase TS+F  as a result of the increase of thermal magnon number and the enhancement  of coupling between the components $\Delta^{\uparrow}$ and $\Delta^{\downarrow}$ of the order parameter. In contrast, in the phonon driven pairing the transition from the superconducting to normal phase on warming is due to smoothing of the step in electron population at the Fermi surface by thermal motion. Such a drastically different behavior of spin wave induced pairing in ferromagnet is the direct consequence of spin conservation at magnon exchange. 
	
	As the anisotropy of the ferromagnet increases, the suppression of the components $\Delta^{\uparrow}$, $\Delta^{\downarrow}$ coupling becomes stronger because of the increase in the magnon energy $\epsilon_s$ 
	and the subsequent drop in the number of thermal magnons. Consequently a bigger $g$ is needed to obtain the same $T_{cl}$, see Fig. 1. 
	
	On heating up the system towards $T_m$ the decrease of the components coupling saturates, while effective electron-electron interaction flatters 
	because the decrease of the spin wave amplitudes, proportional the RPA factor $\langle S_z\rangle$, is compensated by the decrease of magnon frequencies. 
	Approaching $T_m$ we enter into the region of strong magnetic fluctuations and our RPA approximation to find spin correlation functions becomes invalid. 
	To find position of the first order phase transition, $T_{ch}(g)$,  one needs more accurate treatment of the spin correlation functions. We think that $T_{ch}$ probably lays below $T_m$. Note, that the second order phase transition
	between N and TS phase is excluded because there is only one solution for $T_c$  with vanishing order parameter, i.e. the line $T_{cl}(g)$ found previously.
	
	Mineev \cite{Mineev} discussed the effect of magnetic fluctuations near $T_m$ on the coupling parameter of triplet pairing in the framework of the BCS (static) approach to find $T_c$. However, in the framework of a static BCS  approach for spin fluctuation exchange the cutoff frequency which essentially characterizes the dynamics of the system and determines $T_{c}$, cannot be well defined. What is more, we see that for the spin-dependent glue the dependence of the superconducting critical temperature on the coupling parameter {\it is not of the BCS type}. 
	
	Remarkably, the superconducting phase is no longer present for $g<g_c(\kappa)$. We determined the critical value of the parameter $g_c\approx 0.192$. Note, that in our model the parameter $g=2|J_{{\bf k}{\bf k'}}S|^2N(0)/T_m$ essentially represents the fraction  of the exchange RKKY contribution to the Curie temperature.  
	
	The order parameters $\Delta^{\uparrow}$ and $\Delta^{\downarrow}$ depend on the angle $\phi$ in the $x,y$ plane perpendicular to the direction of the magnetization. At $T_{cl}$ this dependence reduces to a $\sin\phi$. Such a dependence may be observed in tunneling measurements between the triplet superconductor and the normal metal.

	Let us consider now the dependence of the order parameter on the frequency.  For $\kappa=0.1$  this is displayed in Figs.~2,3 for dimensionless temperatures 
	$T/T_m=0.02, 0.08, 0.15, 1$. The component  $\Delta^{\downarrow}$ transforms into $\Delta^{\uparrow}$ by means of a magnon emission (positive $\omega$). Those excitations have a gap $\kappa$ at $0<\omega<\kappa$ and at $-2-\kappa<\omega<2$. This gap, as well as the square root anomalies, $\propto 1/ \sqrt{\omega-\kappa}$, at the very edges of the spectrum, are shown in Fig. 4.  The dependence 
	$\Delta^{\downarrow}(\omega)$ perfectly reflects the magnon spectrum as was discussed previously in the case of phonon induced superconductors in Ref.~\onlinecite{SSW}.
	The dependence $\Delta^{\uparrow}(\omega)$ corresponds to the magnon absorption (negative $\omega$) with singularities near the edges of the spectrum at 
	$\omega =\kappa$ and $\omega=2+\kappa$. Anomalies near the edges of the spectrum become sharper at higher temperatures due to induced emission or absorption by thermal magnons. Note, that $\Delta^{\downarrow}(\omega)$  and $\Delta^{\uparrow}(\omega)$  satisfy the symmetry relation
	\begin{eqnarray}
	\Delta^{\downarrow}(\omega+\kappa)=\Delta^{\uparrow}(-\omega-\kappa).
	\end{eqnarray}

	\begin{figure}
		\psfig{figure=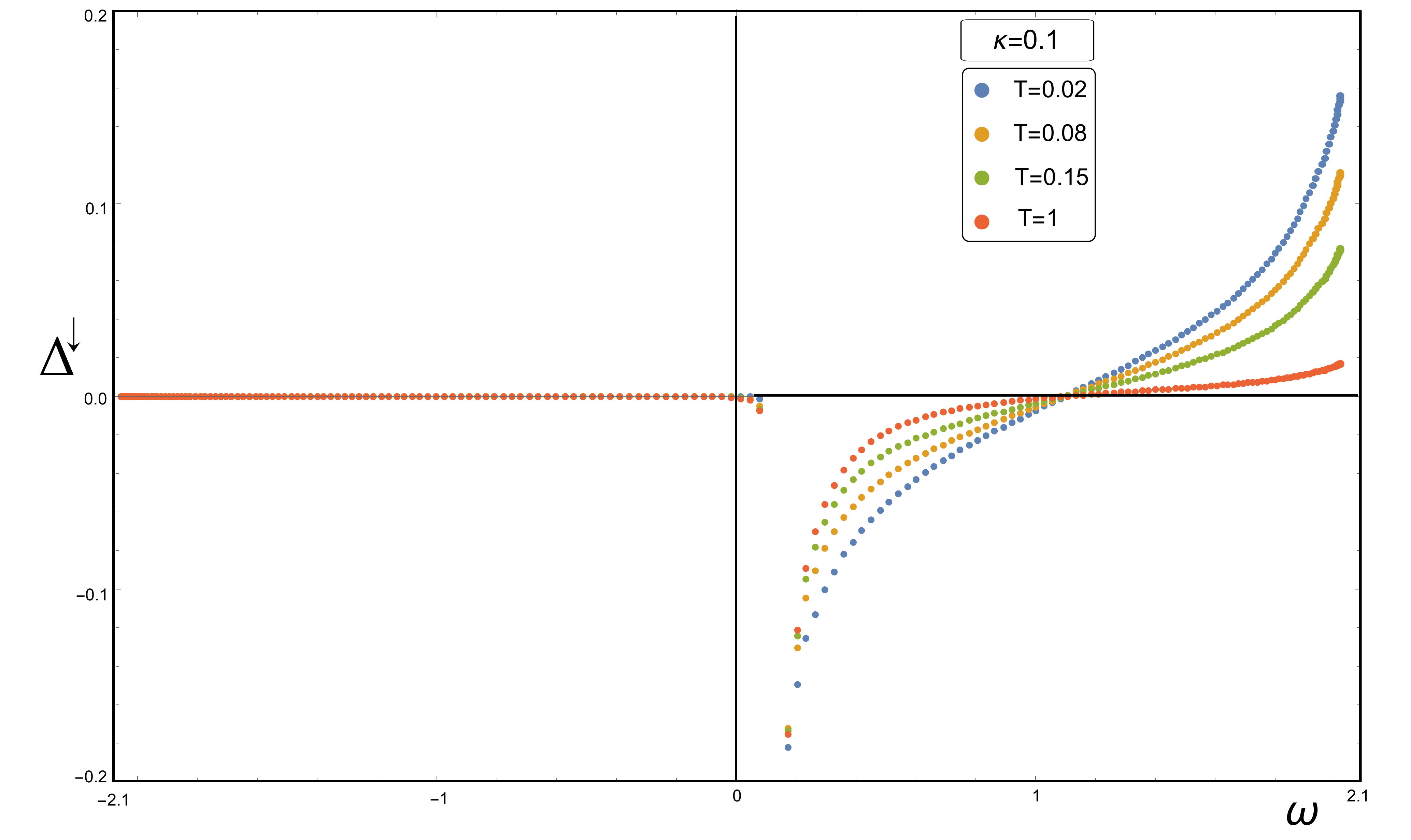,width=\columnwidth}
		\caption{\label{f2}(color online) The frequency dependence of the order parameter $\Delta^{\downarrow}$ in the ferromagnet with magnetic anisotropy $\kappa=0.1$. The dependence at $\kappa<\omega<2+\kappa$ reflects the spectrum of magnons for $\epsilon(\theta)=\kappa+1-\cos\theta$,  where $\theta$ is the polar angle in the $x,y$ plane.}
	\end{figure}
	\begin{figure}
		\psfig{figure=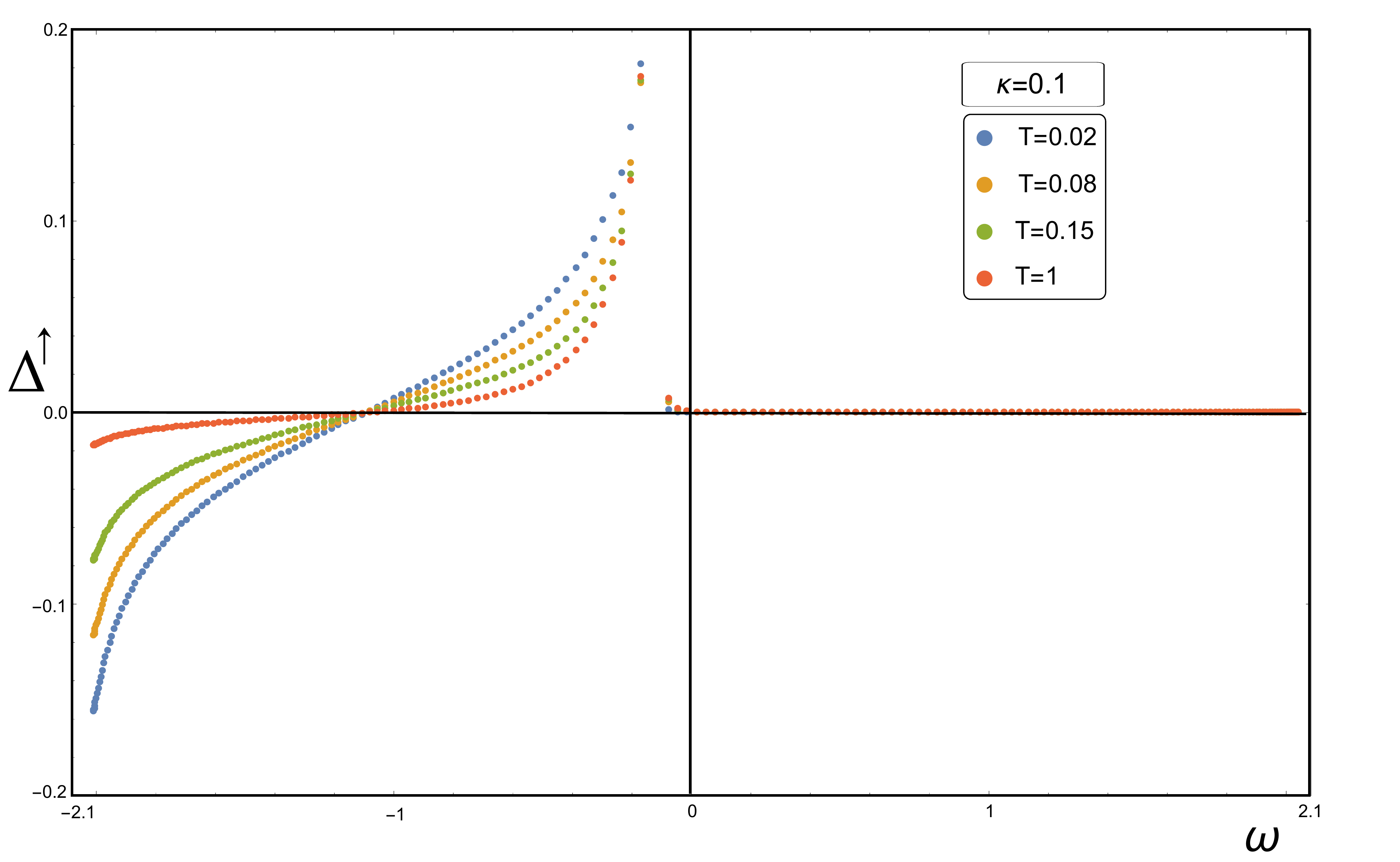,width=\columnwidth}
		\caption{\label{f3}(color online) The frequency dependence of the order parameter $\Delta^{\uparrow}$ in the ferromagnet with magnetic anisotropy $\kappa=0.1$. The dependence at $-2-\kappa<\omega<\kappa<-\kappa$ reflects the spectrum of magnons $\epsilon({\bf q})=\kappa+1-\cos\theta$.}
	\end{figure}
	
	
	Frequency dependence of the order parameter results in similar anomalies in the I-V characteristics of tunneling between the TS and the normal metal, as it was observed in the phonon-induced superconductors (see Ref.~\onlinecite{SSW}).

	Let us compare our results with the magnetic exciton mediated superconductivity attributed to UPd$_2$Al$_3$. \cite{Hale}
	Such an exciton arises from crystal-field-split U4+ levels. It is coupled to delocalized $f$-electrons by electron spin operator $\sigma_z$. Thus
	the electron part of the electron-exciton Hamiltonian  is the same as Eq.~(1),(2), but Eqs. (3),(4) have now the form
	\begin{eqnarray}
	&&{\cal H}=\sum_{\bf q}\omega_{{\bf q}}[\alpha_{{\bf q}}^+\alpha_{{\bf q}}-I\int d{\bf r}\Psi^+_{\alpha}({\bf r})\sigma_{\alpha\beta}^z\Psi_{\beta}({\bf r})]\Phi({\bf r}),\\
	&&\Phi({\bf r})=\sum_{\bf q}\tilde{\lambda_{\bf q}}(\alpha_{\bf q}+\alpha_{-{\bf q}}^+)e^{i{\bf q}{\bf r}},
	\end{eqnarray}
	where $\omega_{\bf q}$ is the dispersion of magnetic exciton, while $\alpha_{\bf q}^+$ and $\alpha_{\bf q}$ are creation and annihilation of magnetic exciton.
	For such a coupling magnetic exciton does not carry spin and acts similar to phonons. Thus the behavior of these triplet superconductors with respect to the temperature is a standard BCS-like. Only triplet superconductors with a spin-carrying mediator (spin wave) differ drastically from superconductors with spin-neutral mediators. 

	\section{Conclusions}
	We used an appropriate spin-electron model Hamiltonian to discuss the spin-wave-mediated triplet pairing in ferromagnets. We found a very particular phase diagram in the plane $(T_{cl},g)$, shown in Fig.~1, for the triplet superconducting state described by the two-component order parameter. The pairing exists only at $g>g_c$ and only in the temperature interval $T_{cl}(g)<T<T_{ch}<T_m$. The low temperature boundary of the TS phase is determined by the conservation of the $z$-component of the spin in the electron-spin wave interaction. It is spin carrying glue and the spin conservation law which make this two-component superconducting phase diagram so different from that of pairing mediated by spin less phonon glue resulting in a single-component order parameter. 
	
	The question now is whether or not this phase diagram is specific for our model Hamiltonian or it is more  generally applicable for any triplet spin wave mediated pairing. 
	The more general non relativistic Hamiltonian to treat the triplet pairing for electrons in $s$-, $d$- or $f$-bands must take into account the strong Coulomb repulsion in the latter narrow bands. In the ferromagnetic phase with the spontaneous magnetization along the $z$-axis, the collective spin wave mode always exists and it is described by the two-electron Green function. 
	The coupling of conduction electrons with this mode results in a triplet pairing as in our model system. This coupling is still governed by the conservation of spin $z$-component. Thus the low temperature boundary of the triplet phase should always exist, as in our model. 
	What is different is that we need to derive the strength of the electron-spin wave coupling $g$ as well as the new $T_{cl}$ and $T_{ch}$ boundaries of triplet phase. Thus, we think that the qualitative behavior of triplet superconductor phase diagram in ferromagnets may be qualitative the same as shown in Fig.~1, but the quantitative behavior may be different in each case.
	
	Experimental data for U-based superconductors show BCS-like phase diagram, i.e. the pairing phase exists at all temperatures below $T_c$.
	Thus we are inclined to think that their superconducting phase is not a triplet state of spin exchange mechanism of pairing. It may be still phonon based pairing in the presence of strong spin-orbit interaction which suppresses destructive effect of exchange field for singlet pairing or it may be some other hypothetical mechanism 
	resulting in static effective attraction of electrons.
	
	Numerical calculations were performed on the High-Performance Computing Center (NPAD) supercomputer at the Federal University of Rio Grande do Norte (Natal, Brazil).
	The authors acknowledge useful conversations with V. Mineev, V. Kogan and D. Khomskii.


\begin{thebibliography} {10}
		\bibitem{Eliashberg}G.M. Eliashberg, Zh. Eksperim. i Teor. Fiz. {\bf 38}, 966 (1960);
		{\bf 39}, 1437 (1960) [Soviet Phys.  JETP {\bf 11}, 696 (1960); {\bf 12}, 1000 (1961)~
		\bibitem{AGD}A.A. Abrikosov, L.P. Gorkov, I.E. Dzyaloshinski, {\it Methods of Quantum Field Theory in Statistical Physics}, Oxford press, 1965.
		\bibitem{SSW}D.J. Scalapino, J.R. Schrieffer, J.W Wilkins, Phys. Rev. {\bf 148}, 263 (1966).
		\bibitem{Scalapino}D.J. Scalapino, in {\it Superconductivity}, v.1, Edited by R.D. Parks, 1969, p. 449.
		\bibitem{McMillan}W.L. McMillan, Phys. Rev. {\bf 167} 331 (1968)
		\bibitem{Aoki1}Aoki D, Hardi F, Miyake A, Taufour V, Matsuda T D, Flouquet J Compt. Rend. Physique {\bf 12} 573 (2011)
		\bibitem{Aoki2} Aoki D and Flouquet J J. Phys. Soc. Jpn. {\bf 81} 011003 (2012).
		\bibitem{Mineev}V.P.~Mineev, Usp. Fizich. Nauk {\bf 187}, 129 (2017) [Phys. Usp. {\bf 60}, 121 (2017)], arXiv:1605.07319v1
                \bibitem{Aoki3}D. Aoki and J.J. Flouquet. Phys. Soc. Jpn.{\bf 83} 061011 (2014).
		\bibitem{Fay}D. Fay, J. Appel, Phys. Rev. B, {\bf 22}, 3173.
		\bibitem{Hale}P. McHale, P. Fulde, P. Thalmeier. Phys. Rev. B {\bf 70} 014513 (2004)
		\bibitem{Hattori} K. Hattori, H. Tsunesugu, Phys. Rev. B {\bf 87}, 064501 (2013).
		\bibitem{Tyab}S. V. Tjablikov, {\it Methods in the quantum theory of magnetism}
		(Plenum Press, New York, 1967).
		\bibitem{Moriya}T. Moriya, Phys. Rev. {\bf 120}, 91 (1960).	
		\bibitem{Kirkpatric}T. R. Kirkpatrick, D. Belitz, Phys. Rev. B,  {\bf 67} 024515 (2003).
		.\bibitem{Woolsey}R.B. Woosley, R.M. White, Phys. Rev. B {\bf 1}, 4474 (1970).
		\bibitem{GW}G. W. Recktenwald, Numerical Methods With MATLAB: Implementations and Applications
		(Prentice Hall, 2000).
	\end{thebibliography}
	\end{document}